# Mechanical consequences of dynamically loaded NiTi wires under typical actuator conditions in rehabilitation and neuroscience

Umut D. Çakmak[a*], Zoltán Major[a] and Michael Fischlschweiger[b*]

[a]*Johannes Kepler University Linz, IPPE, Altenbergerstrasse 69, 4040 Linz, Austria*

[b]*OTTRONIC Technology Laboratory, OTTRONIC GmbH, Villenstrasse 10, 8740 Zeltweg, Austria*

umut.cakmak@jku.at

m.fischlschweiger@ottronic.com

## Abstract

In the field of rehabilitation and neuroscience shape memory alloys play a crucial role as lightweight actuators. Devices are exploiting the shape memory effect by transforming heat into mechanical work. In rehabilitation applications, dynamic loading of the respective device occurs, which in turn influences the mechanical consequences of the phase transforming alloy. Hence in this work, dynamic thermomechanical material behavior of temperature triggered phase transforming NiTi shape memory alloy (SMA) wires with different chemical compositions and geometries is experimentally investigated. Storage modulus and mechanical loss factor of NiTi alloys at different temperatures and loading frequencies are analyzed under force controlled conditions. Counterintuitive storage modulus and loss factor dependent trends regarding the loading frequency dependency of the mechanical properties on the materials' composition and geometry are hence obtained. It could be revealed that loss factors show a pronounced loading frequency dependency, whereas the storage modulus was not affected. It is shown that force controlled conditions lead to a lower storage modulus than expected. Further it turned out that a simple empirical relation can capture the characteristic temperature dependency of the storage modulus, which is an important input relation for modeling the rehabilitation device behavior under different dynamic and temperature loading conditions, taking directly into account the material behavior of the shape memory alloy.

## Keywords

# 1   Introduction

To extend the duration of therapeutic sessions, robotic systems are becoming a very efficient tool in the field of rehabilitation and neuroscience (Pittaccio and Viscuso 2012). In rehabilitation, robotic tools enable repetitive and repeatable mobilization of the whole limb (Hesse 2008, Pittaccio and Viscuso 2012) or joint by joint (Krebs et al. 2001). Managing the motion of effectors by robotic systems gains information of the interconnection between activities of peripheral segments and brain structures, hence this is a promising approach in neuroscience (Boonstra et al. 2005, Onishi et al. 2010, Hollnagl et al. 2011, Pittaccio and Viscuso 2012). Consequently, high interest is devoted to portable lightweight devices, where actuation is based on shape memory alloys.

The understanding of shape memory alloys (SMA) as a class of metals with the capability of changing their shape due to applied thermal, mechanical or magnetic fields has been studied in the last decades (Schetky 1979, Bhattacharya et al. 2004, Zheng et al. 2004, Bhattacharya 2003, Aaltio et al. 2008, Otsuka et al. 2011, Oberaigner and Fischlschweiger 2011, Yastrebov et al. 2014, Mahtabi et al. 2015). The physical mechanism behind the shape memory effect or the material's superelasticity is a displacive solid to solid phase transformation, where due to an externally applied field a spontaneous change of the crystallographic lattice occurs, hence activating a change of the macroscopic mechanical properties (Basu et al. 1999, Humbeeck, 2003, Shaw et al. 2003, Bhattacharya et al. 2004, Waitz et al. 2007, Fischlschweiger and Oberaigner 2012, Yastrebov et al. 2014).

Depending on the alloy's composition, the thermomechanical characteristics can be varied. Nowadays, owing to numerous research efforts, a better understood shape memory effect of, e.g., Cu-Zn-Al, Cu-Al-Ni and Ni-Ti alloys, allows more sophisticated applications in different fields (Wei et al. 1998a, Levitas and Preston 2002, Neuking et al. 2005, Idesman et al. 2008, Hartl and Lagoudas 2008). For medical applications NiTi alloys are typically be used as implants and as superelastic elements and actuators in the field of rehabilitation and neuroscience (Pittaccio and Viscuso 2012).

However, for NiTi alloys, there are still open issues and unsolved aspects regarding the development of alloys with specific and tailor-made properties for applications concerning temperature and dynamic mechanical loading-dependent material characteristics (Duerig, 2006). These issues have not been fully investigated in the past (Duerig 2006, Roy et al. 2008). A high number of commercial products made of SMA are produced in the form of wires (Wei et al. 1998a and 1998b) which are further operating as actuators and damping elements in devices. Especially, for a better understanding of the materials' damping capacity in connection with variations of effective modulus, fundamental knowledge is required (Duerig, 2006) to support modelling and prototyping of devices. An important step in the design of a SAM-based rehabilitation actuator is a careful analysis of biomechanical boundary conditions, i.e. resisting loads (Pittaccio and Viscuso 2012). This leads to the constraint that force controlled boundary conditions have a high relevance in rehabilitation actuators, whereas in literature information about the material behavior under these conditions can rarely be found.

Additionally, SMAs are deploying their damping capacity dependent on the external thermomechanical loading. For rehabilitation and neuroscience actuators the shape memory effect is triggered by temperature induced phase transformation (Pittaccio and Viscuso 2012) under a respective stress level which is of course below the stress induced transformation level. Contrary to temperature induced phase changes, stress triggered phase transformations in SMAs exhibit higher damping capacities and consequently they have been more focused on in materials science in the past with respect to dynamic amplitude and frequency effects (Piedboeuf et al. 1998, Young et al. 2013).

Therefore, in this work the focus lies on the understanding of dynamic mechanical properties of NiTi-SMA wires, under controlled temperature and force fields for various mechanical loading frequencies in connection with effective modulus change, especially in the temperature triggered transformation zone. Earlier experimental investigations on NiTi alloys revealed properties influencing the materials' damping capacity in the transformation zone (Sugimoto and Nakaniwa 2000, Debdutta et al. 2008). It was shown that material softening takes place during the transformation referred to the observation that the modulus of the SMA was lower in the transition region than the moduli of austenite and martensite in the pure phase state. The softening effect could be also explained and predicted theoretically by using statistical mechanics based modeling strategies developed by Oberaigner and Fischlschweiger (2011). Further Yastrebov et al. (2014) modeled this phenomenon with Lattice-Monte-Carlo approach. In addition, Wang and Sehitoglu (2014) highlighted the dramatic difference of the macroscopic modulus influenced by the multivariante state of martensite. Especially the selection of martensitic variants is dependent on external applied stress fields and is sensitive to tension and compression. Hence, moduli of NiTi alloys are different in tension and in compression (Wang and Sehitoglu 2014). Here we analyze the macroscopic behavior of NiTi alloys under dynamic thermomechanical loading under force controlled conditions to gain insights into the loading-dependent damping and modulus behavior of SMA wires with different geometries and materials' compositions. This information is relevant for selecting NiTi-wires as actuator and damping elements in rehabilitation and neuroscience devices. We present the experimental methodology to characterize the dynamic thermomechanical behavior of NiTi alloys. Special emphasis is given to the analysis of the stiffness and the mechanical transient properties, i.e., loss factor. Furthermore, we seek to find a simple empirical relation of the temperature dependent effective modulus behavior. This is particularly important to take into account the respective SMA material properties in modeling rehabilitation and neuroscience devices.

## 2 Experimental section

### 2.1 Materials

The NiTi alloys under investigation were commercial products of Memry Corp. (USA). Table 1 lists the compositions, the diameters of the round wires, and the surface conditions of the alloys. All wires were straight annealed (superelastic) by the manufacturer and, with the exception of alloy M5, the materials' surfaces were oxidized. Quasistatic tensile tests (Electromechanical actuator, Testbench, Bose Corp.,

ElectroForce Systems Group, MN, US) at room temperature (22°C) were performed in order to determine the Young's moduli of the studied alloys. The moduli are also listed in Table 1.

All alloys show R-phase transformation behavior and transform from B2 cubic austenite sequentially via rhombohedral R-Phase to B19' monoclinic martensite. Besides, functional fatigue properties it is typically for alloys with R-phase transformation behavior that small transformation strains are achieved (Sittner et al. 2006).

The alloys were selected according to their practical use as wires with differences in alloy composition (stiffness), geometry and surface finishing, in order to analyze the respective behavior under dynamic thermomechanical loadings. As a remark, oxidized surface finishing is particularly relevant in cases, where the device has to fulfill sterilization requirements.

Table 1 Specifications of the investigated alloys.

| Alloy | Ni % | Ti % | Diameter μm | Surface finish -- | Young's Modulus $E$ @ 22°C GPa |
|-------|------|------|-------------|-------------------|-------------------------------|
| M1    | 55.60 | 44.40 | 1000 | Oxide      | 13 |
| M2    | 55.34 | 44.66 | 320  | Oxide      | 30 |
| M3    | 55.77 | 44.23 | 323  | Oxide      | 33 |
| M4    | 54.80 | 45.20 | 305  | Oxide      | 24 |
| M5    | 55.34 | 44.66 | 760  | Oxide free | 15 |

## 2.2 Characterization Methodology

On principle, there exist different experimental methodologies for studying the modulus behavior of NiTi alloys. This spans from typical quasistatic mechanical tests, ultrasound methods up to thermomechanical and dynamic thermomechanical methods (e.g., Sadjadpour et al. 2015, Daly et al. 2007, Sittner et al. 2006, Van Humbeeck 2001, Shaw and Kyriakides 1995). It is already known that pure phase elastic properties of NiTi alloys with R-phase transformation obtained by ultrasonic transmission measurements agree well with those obtained by mechanical tests. However, a disagreement appears in the R-phase transformation region (Sittner et al. 2006). In ultrasonic measurements elastic constants are determined by evaluating the propagation speed of longitudinal and transversal ultrasonic waves. As it is documented in the work by Sittner et al. 2006, the wave speed is additionally affected by microstructure, which results in questionable experimental results in the temperature ranges, where phase mixtures exist. In this work the focus is on frequency dependent dynamic modulus behavior especially within the transformation zone of NiTi alloys showing R-phase transformation behavior under force controlled conditions.

Here, the characterization methods included differential scanning calorimetry (DSC) for determining the character of phase transformation and dynamic thermomechanical analysis (DTMA) under tensile loading conditions. DSC is further used to investigate the impact of thermal rate effects on shifting the phase transformation behavior with respect to heating and cooling. So, the impact of occurring thermal rate changes in the temperature control chamber of DTMA on phase transformation for ensuring a correct interpretation of DTMA measurements. An equivalent DSC experimental procedure – as reported by Zheng et al. (2003) – was conducted using a Mettler Toledo DSC 822 with a DDK FRS5 sensor, an intercooler TC100MT-NR and a gas controller GC 200. The heating and cooling rate was 10 K/min for the standard measurements. This corresponds to the typical heating rate in the temperature control chamber. Herein, a heating/cooling rate-dependent DSC measurement was performed for one of the investigated alloys to analyze the kinetic effect regarding the influence on the transformation temperatures. Frequency sweeps were performed under isothermal conditions, however, rates for setting the desired temperature can be varied and may influence the specimen's thermomechanical behavior. For determining the impact of thermal rate effects on the thermomechanical responsiveness of the material, heating and cooling rates of 0.1 K/min and 1 K/min were additionally examined.

DTMA (TestBench, Bose Corp., ElectroForce Systems Group, MN, US) is used to study the frequency dependence of modulus behavior of samples given in Table 1, working out especially the discrepancy of storage and loss factor for the respective sample. During the DTMA experiments, a force-controlled sinusoidal excitation was applied under isothermal conditions. A detailed description of the measurement setup and procedure can be found in the work by (Cakmak and Major 2014). The excitation mean level was 25 MPa, and the dynamic amplitude was 5 MPa during the frequency sweep from 1 Hz to 21.5 Hz at temperatures changing in 10 K (5 K) increments from 353.15 K to 243.15 K. The heating and cooling rate between the isothermal steps was in a range of (10 +/- 1) K/min. At each temperature step, the initial length at thermal equilibrium was considered. The storage modulus $E'$ and the loss factor $\tan\delta = E''/E'$ were determined from each isothermal measurement.

## 3 Results and Discussion

### 3.1 Differential Scanning Calorimetry (DSC)

Table 2 summarizes the thermal characteristic transition temperatures from DSC measurements. All alloys revealed an R-phase and the corresponding start and finish transition temperatures are listed in Table 2.

Table 2 Thermal characteristics of the investigated alloys. Austenite start and finish transition temperatures ($A_S$ and $A_F$), R (rhombohedral or R-Phase) start and finish temperatures ($R_S$ and $R_F$) and martensite start and finish temperatures ($M_S$ and $M_F$).

| Alloy | $T$-rate (K/min) heating/cooling | $A_S$ (K) | $A_F$ (K) | $R_S$ (K) | $R_F$ (K) | $M_S$ (K) | $M_F$ (K) |
|---|---|---|---|---|---|---|---|
| M1 | 10 | 327.86 | 339.24 | 329.84 | 309.89 | 272.86 | 215.08 |
| M2 | 10 | 324.30 | 341.10 | 327.73 | 301.16 | 298.53 | 245.48 |
| M3 | 10 | 290.99 | 326.26 | 323.91 | 283.45 | -- | -- |
| M4 | 10 | 355.61 | 362.82 | 336.65 | 332.19 | 319.28 | 309.92 |
| M5-10 | 10 | 317.19 | 328.46 | 313.30 | 307.58 | 266.61 | 258.49 |
| M5-1 | 1.0 | 316.24 | 323.15 | 313.71 | 309.84 | 267.82 | 264.33 |
| M5-0.1 | 0.1 | 316.12 | 323.05 | 314.43 | 309.54 | 268.89 | 265.93 |

The kinetic effect during the DSC measurement was investigated for alloy M5 (see FIG. 1). With an increasing heating rate, the austenite transition temperatures (AS and AF) also increased slightly. However, R-phase and martensite transition temperatures decreased slightly with a higher cooling rate. Generally, the kinetic effects on the transformation temperatures were relatively low and in accordance with the conclusion of Nurveren et al. (2008). It can be concluded that the heating/cooling rate for setting the required temperature for DTMA experiments will not affect the behavior substantially of the materials under investigation. Therefore temperature rate effects in the DTMA with the current parameters can be excluded, this holds true for peak, as well as for start and finish temperatures.

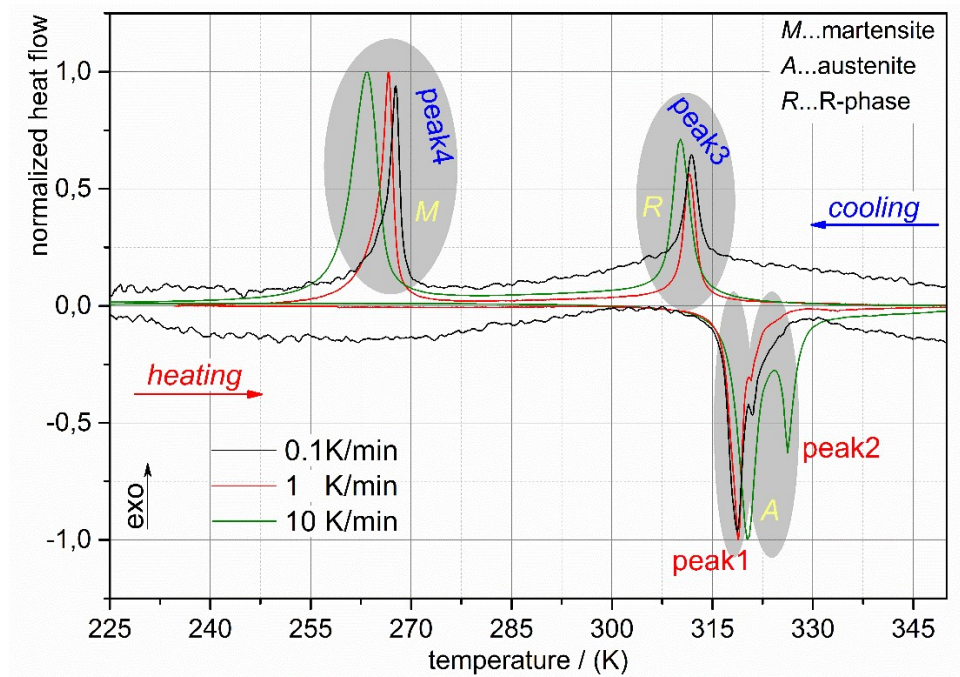

FIG. 1 DSC results of the M5 alloy showing the temperature memory effect for three different heating/cooling rates. For the sake of visualization and comparability the heat flow is normalized.

## 3.2 Dynamic Thermomechanical Analysis (DTMA)

In the following diagram (see FIG. 2), the examined $E'$ results at the investigated temperatures are illustrated for all NiTi alloys. First of all, the experimental data points indicate that the materials' storage modulus is, as earlier discussed and expected, strongly temperature dependent. Generally, the course of the data points is similar to earlier reported results in the literature from model predictions and experimental findings (Fischlschweiger and Oberaigner 2011, Fischlschweiger et al. 2012, Sugimoto and Nakaniwa 2000, Roy et al. 2008, Yastrebov et al. 2014) and is characteristic for NiTi alloys with different chemical compositions. The temperature determines the actual state of the microstructure regarding the content of austenite, martensite and the intermediate phase (R-phase).

Some of the alloys reveal a very narrow transition region, as is the case for alloys M1, M3 and M5. Alloy M4 shows only a part of the transition region within the investigated temperature range. This is due to the alloy's formulation with transition temperatures higher than those of the others. Alloys M2 and M5 are basically the same formulation, but with oxide and oxide-free surfaces, respectively. The comparison of M2 with M5 reveals that the temperature-dependent storage modulus characteristic is also influenced by the surface treatment. M2 exhibits, generally, higher moduli values and, between 260K and 320K, a rather constant modulus (~28GPa). In contrast, the transition region of alloy M5 starts at 320K and ends at 313K. Comparing M5 with M2, the storage modulus within the transformation region of M5, where phase mixture exist, is in contrast to M2 not constant, rather the storage modulus changes continuously. This information is of particular interest, because it shows, that an oxide layer on the NiTi wire increases the dynamics storage modulus with the factor 2 and is hence a remarkable parameter for controlling the dynamic stiffness of NiTi wires and consequently of the actuator device. Due to the oxide layer a more constant modulus behavior within the transformation zone exists in a broader temperature interval, this has a high impact on material's selection for the respective use case.

In addition, the dynamic storage Young's moduli at room temperature are lower than those of the quasistatic tensile tests (cf. Table 1). This is in analogy to earlier investigations (Miyazaki et al. 1986, Schmoller and Bausch 2013), where cyclic softening was observed and analyzed.

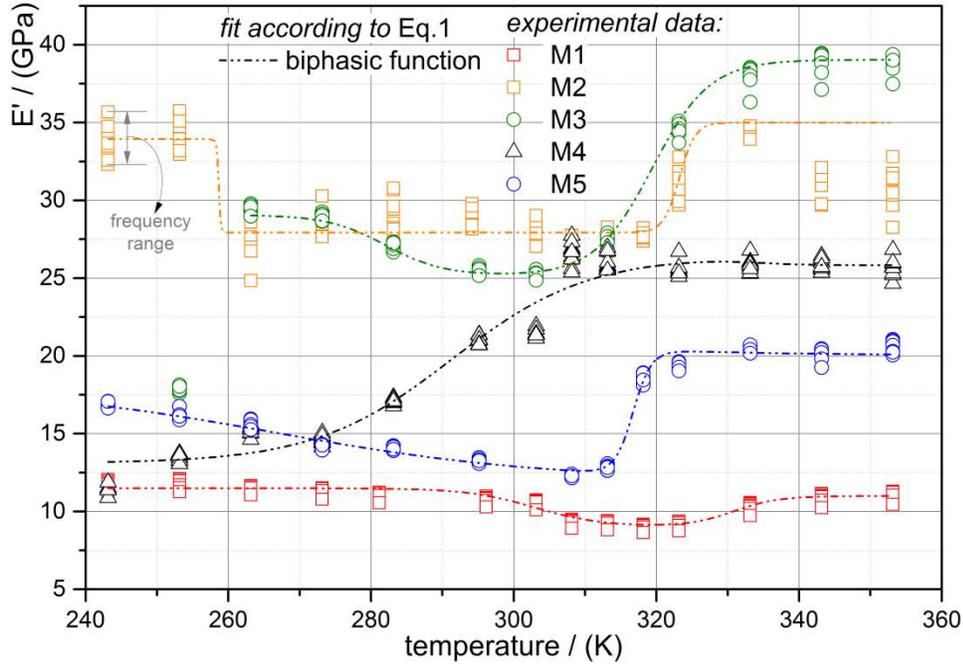

FIG. 2 Comparison of the temperature dependent moduli of all alloys within the examined frequency range of 1 to 21.5 Hz. Dashed lines are the modeled curves according to Eq.1.

Moreover, from the DTMA measurements, the mechanical loss factor $\tan\delta$ was evaluated, and the temperature- and excitation frequency-dependent characteristics of the alloys can be seen in FIG. 3. To facilitate the observability of the temperature-dependent loss-factor trend, the average $\tan\delta$ over the examined frequency range is shown in the diagram. In addition, the maximum and minimum $\tan\delta$ (dashed) lines are illustrated, and arrows indicate the observed $\tan\delta$ within the investigated frequency range. Generally, the evaluated $\tan\delta$ values are reasonable and are within the limit of the mechanical loss factors of steel (0.002 to 0.01) (cf. de Silva 2000). A counterintuitive frequency dependency is observable by comparing the mechanical loss factor with the earlier presented storage moduli data (see FIG. 2). The moduli of the alloys are clearly frequency independent, while the loss factors alter in discrepancy to the former with varying excitation frequencies. The lower the loading frequency, the higher the loss factor; whereas the course of the temperature-dependent curve is equivalent. This discrepancy between the mechanical transient and storage behaviors of NiTi alloys affects the performance of devices during dynamic loading. It mainly affects the damping behavior, while the resonance frequency of the structure is unaffected. On the other hand, the difference between the investigated alloys' $\tan\delta$ characteristics was not as distinctive as the modulus characteristics; less variation between the alloys was observed. Alloys M3 and M5 revealed a similar characteristic progression and showed an increase from low temperatures up to a peak followed by a decline to a minimum; whereas the other alloys were, in principle constant between 270 K and 320 K.

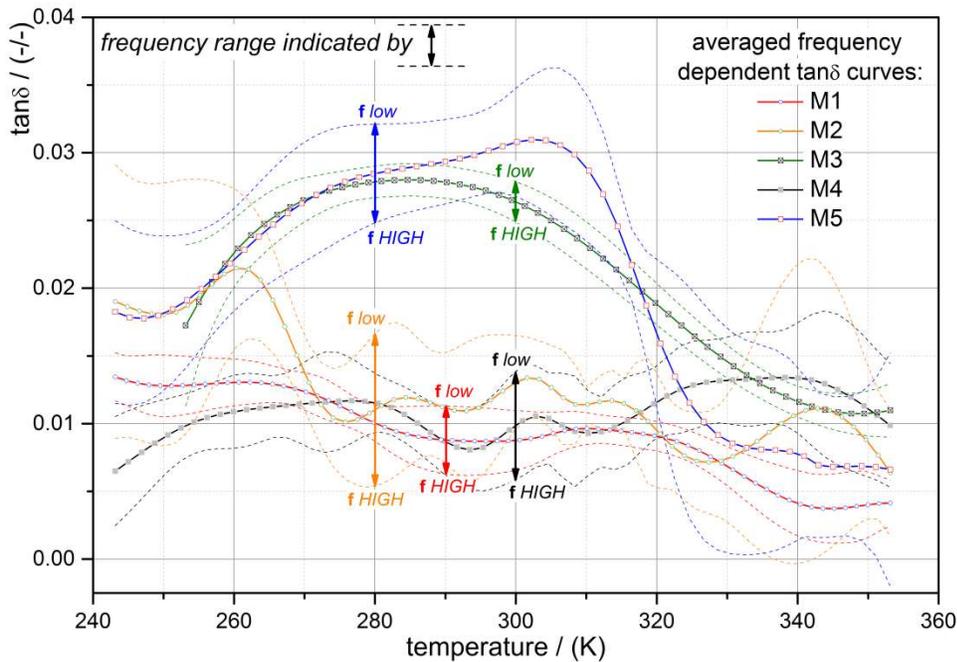

FIG. 3 Comparison of the temperature dependent mechanical loss factor $\tan\delta$ (determined from DTMA experiments). Full lines represent the over the frequency range averaged curves and the dashed lines show the respective high and low frequency lines.

The discrepancy of frequency dependence of storage modulus and loss factor can be physically explained, by the fact, that the viscos part of all NiTi alloys investigated in this study is almost negligible and therefore a frequency independent storage modulus behavior was observable. The modulus behavior itself is, however, strongly dependent on respective microstructure realized by different chemical compositions and by oxide surface layers of the wires. The measured storage modulus behavior is lower than typically obtained modulus by quasistatic tests and ultrasonic measurements. The reason lies in the dynamic measurement procedure carried out under controlled stress and not strain boundary conditions. Hence, internal strains are developing due to the cyclic loading to higher values ending in a plateau by constant external force conditions. Consequently, lower storage moduli are obtained with respect to for example quasistatic measurement conditions. The frequency dependent loss factors occur due to internal friction phenomena, strongly pronounced especially in the phase mixtures regions. Of particular interest, is that internal friction seems to be similar within the investigated NiTi alloy ending in a similar, however frequency dependent damping capacity of the alloys.

Preparing the experimental storage modulus differences according to chemical composition and surface treatment for designing actuator elements and taking them into account in model calculation, in the next paragraph an empirical description of storage modulus behavior is presented.

From the storage modulus curves, it could be figured out that the modulus behavior is temperature $T$ dependent and frequency $f$ independent, it turned out that $E(T,f) = E(T)$. $E(T)$ can be modeled by the biphasic function type:

$$E(T) = E_{PT} + \frac{E_A - E_{PT}}{1 + e^{(T - T_{0_1})h_1}} + \frac{E_M - E_{PT}}{1 + e^{(T_{0_2} - T)h_2}} \tag{1}$$

where $E_{PT}$ is the soft modul point, meaning the modulus point with the lowest value in the transformation zone, $E_A$ is the modulus of the austenite dominant state of the alloy, $E_M$ is the respective modulus of the martensite dominant state, the temperature, $T_{0_1}$ and $T_{0_2}$ are the temperatures related to the inflection points of the curves, and $h_1$ and $h_2$ are dimensionless constants of the exponential functions. For the alloy M5, the modeled curves are illustrated in FIG. 4. At each measured excitation frequency, the parameters for the function in Eq.1 were determined. In FIG. 5, the moduli parameters of the biphasic function are shown. Up to 21.5 Hz, the determined parameters are almost constant, and this confirms that the moduli of the alloys are frequency independent.

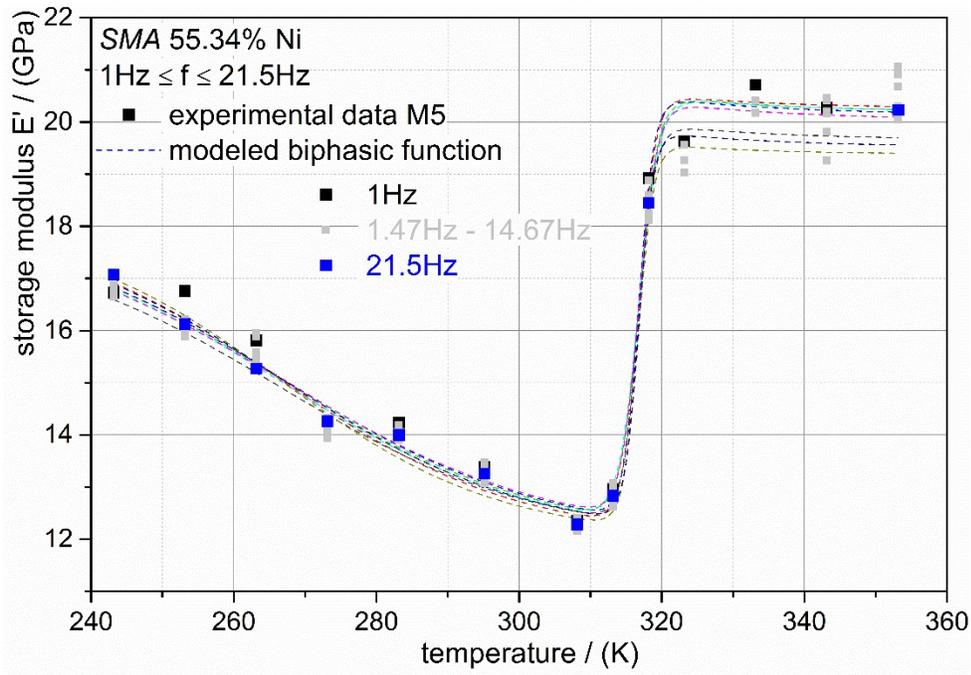

FIG. 4 DTMA results of the alloy M5 showing the temperature dependent storage moduli. Data points represent the experimental data in the investigated frequency range; Dashed lines are the fitted curves of the data according to Eq.1.

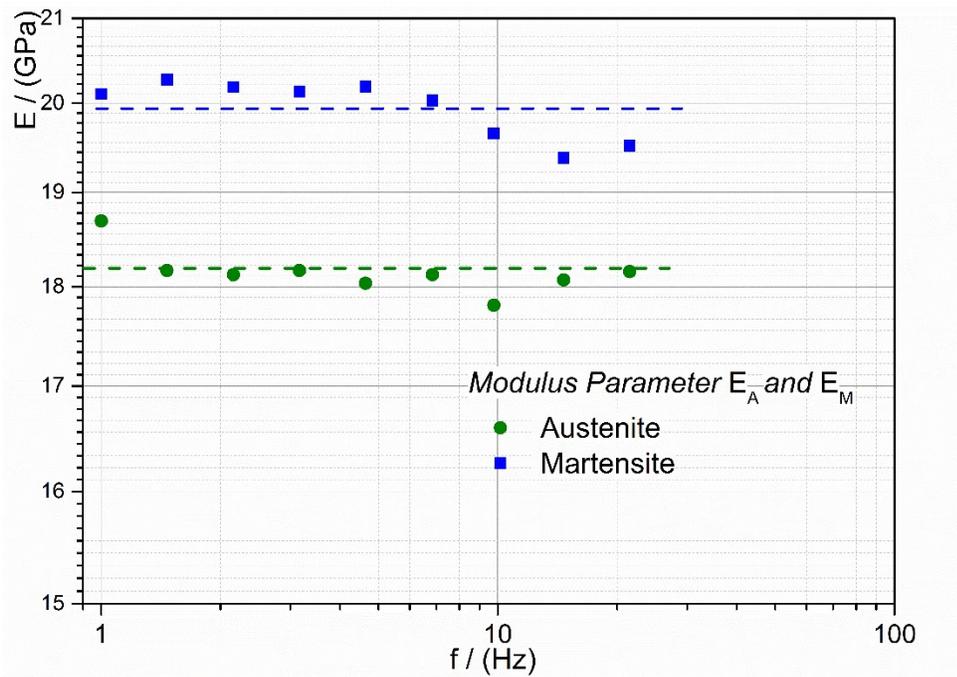

FIG. 5 The frequency dependent modulus parameter of the biphasic function (Eq.1) for the alloy M5.

The capability of the proposed model (Eq.1) is shown in FIG. 2. The DTMA data are sufficiently adjusted by the biphasic function, indicating the applicability of the function. From the experimental data determined model parameters are shown in Table 3. The difference of the alloy composition (M1 to M5) leads also to a variation of the mechanical behavior, in which the moduli related to the phase transition $E_{PT}$ (R-phase) are lower than those of the other phases ($E_A$ and $E_M$). All model parameters (see Table 3) are constant in terms of excitation frequency. The proposed mathematical function is capable of describing the different temperature-dependent modulus characteristics of the alloys and justifies the applicability to NiTi alloys.

Table 3 Parameters for the proposed model in Eq.1 for all investigated NiTi alloys.

| Alloy | $E_{PT}$ | $E_A$ | $E_M$ | $T_{0_1}$ | $T_{0_2}$ | $h_1$ | $h_2$ |
|---|---|---|---|---|---|---|---|
|  | MPa | MPa | MPa | K | K | -- | -- |
| M1 | 9025 | 11492 | 10994 | 302.93 | 330.83 | 0.21 | 0.32 |
| M2 | 27922 | 33941 | 34996 | 258.66 | 323.22 | 7.01 | 0.90 |
| M3 | 25152 | 29070 | 39038 | 282.05 | 319.44 | 0.24 | 0.25 |
| M4 | 12500 | 13098 | 25850 | 333,82 | 291.28 | 0.24 | 0.11 |
| M5 | 12020 | 17810 | 19657 | 266.55 | 315.75 | 0.06 | 0.87 |

## 4  Summary and Conclusions

A series of dynamic thermomechanical analyses (DTMA) were conducted to study the frequency dependency of the mechanical behavior of NiTi alloys with different chemical compositions and wire dimensions. The phase transformation is induced thermally, where the applied dynamic stresses are in a range below the level of stress induced transformation. This setup becomes highly relevant for the understanding of SMA wires which are essential actuator elements in the field of rehabilitation and neuroscience. The experiments were performed force controlled, taking into account practical actuator conditions from rehabilitation and neuroscience application. Based on the above stated conditions, the storage modulus and the mechanical loss factor were determined and analyzed over various frequencies for typical temperature courses. All alloy compositions and geometries revealed an almost frequency-independent storage modulus, while the mechanical loss factor showed a pronounced dependency on the excitation frequency. This discrepancy is of particular interest when dynamically loaded devices are considered. The damping capability is altered by the external mechanical loading excitation and has consequences in the overall performance of the material. At the same time, the stiffness of the material remains unchanged and hence the resonance frequency of the structure. Further it is shown that the dynamic modulus for force controlled loading conditions as it is the case in the application field, the modulus is considerably lower than the quasistatic measured modulus. This should be taken into account in the design of generally dynamically loaded rehabilitation actuators which are bound to controlled force conditions. To exclude any kinetic effect during the temperature ramp in DTMA between the investigated temperatures and during the frequency sweep protocol, differential scanning calorimetry measurements were performed for different cooling rates and hence thermal properties of several alloy compositions were characterized.

It can be concluded that the NiTi alloys under investigation show almost no kinetic effect to heating/cooling rate; however they exhibit kinetic effects to mechanical loading rate (frequency). Only the damping behavior ($\tan\delta$) is affected by the mechanical rate; the stiffness ($E'$) is frequency independent.

Additionally, a simple empirical function describing the temperature dependent storage modulus behavior is proposed. This enables to directly consider temperature dependent dynamic storage modulus behavior for the selected materials as relevant material input property for structural actuator design and modelling.

# 5 Acknowledgement

The research for this paper was performed in part within the framework of the Regio13 project, with the contribution of the Linz Center of Mechatronics GmbH (LCM) in Austria. Regio13 is funded by the European Regional Development Fund (ERDF) and the State Government of Upper Austria (Wi-225867-2012).

# 6 References

Aalito, I., Mohanchandra, K.P., Hezcko, O., Lahelin, M., Ge, Y., Carman, G.P., Söderberg, O., Löfgren, B., Seppälä, J., and Hannula, S.-P., "Temperature dependence of mechanical damping in Ni-Mn-Ga austenite and non-modulated martensite", Scripta Materialia, **59**, 550-553 (2008).

Basu, B., Donzel, L., Van Humbeeck, J., Vleugels, J., Schaller, R., and Van Der Biest, O., "Thermal expansion and damping characteristics of Y-TZP", Scripta Materialia, **40**, 759-765 (1999).

Bhattacharya, K., "Microstructure of martensite, why it forms and how it gives rise to the shape-memory effect", Oxford University Press (2003).

Bhattacharya, K., Contia, S., Zanzotto, G., and Zimmer, J., "Crystal symmetry and reversibility of martensitic transformation", Nature, **428** 55-59 (2004).

Boonstra, T., Clairbois, H., Daffertshofer, A., Verbunt, J., van Dijk, B., Beek, P., "MEG-compatible force sensor" Journal of Neuroscience Methods, **144**, 193-196 (2005).

Çakmak, U. D. and Major, Z., "Experimental Thermomechanical Analysis of Elastomers Under Uni-and Biaxial Tensile Stress State",Experimental Mechanics, **54**(4), 653-663 (2014).

Daly, S., Ravichandran, G., and Bhattacharya K., "Stress-induced martensitic phase transformation in thin sheets of Nitinol." Acta Materialia, **55** (10), 3593-3600 (2007).

De Silva, C. W., "Vibration: fundamentals and practice", CRC Press, Boca Raton, FL, US (2006).

Duerig, T.W., "Some unsolved aspects in Nitinol",Material Science and Engineering A, **438**, 69-74 (2006).

Fischlschweiger, M., Cailletaud, G., and Antretter, T., "A mean-field model for transformation induced plasticity including backstress effects for non-proportional loadings", International Journal of Plasticity, **37**, 53-71 (2012).

Fischlschweiger, M., Oberaigner, E.A., "Kinetics and rates of martensitic phase transformation based on statistical physics", Computational Materials Science, **52**, 189-192 (2012).

Hartl, D.J., and Lagoudas, D.C., "Thermomechanical characterization of shape memory alloy materials" Springer, US (2008).

Hesse, S., "Treadmill training with partial body weight support after stroke: a review" Neuro Rehabilitation, **23**, 55-65 (2008).

Hollnagel, C., Brügger, M., Vallery, H., Wolf, P., Dietz, V., Kollias, S., Riener, R., "Brain activity during stepping: a novel MRI-compatible device" Journal of Neuroscience Methods, **201**, 124-130 (2011).

Idesman, A.V., Cho, J.-Y., and Levitas, V.I., "Finite element modeling of dynamics of martensitic phase transitions" Applied Physics Letters, **93**, 043102 (2008).

Krebs, H., Rossi, S., Kim, S., Artemiadis, P., Williams, D., Castelli, E., Cappa, P., "Pediatric anklebot" IEEE Int. Conf. Rehabil. Robot 1-5 (2011).

Levitas, V.I., and Preston, D.L., "Three-dimensional Landau theory for multivariant stress-induced martensitic phase transformations. I. Austenite ↔ martensite" Physical Review B, **66**, 134206 (2002).

Mahtabi, M., Shamsaei, N., Mitchell, M., "Fatigue of Nitinol: the State-of-the-Art and Ongoing Challenges" Journal of the Mechanical Behavior of Biomedical Materials, **50**, 228-254 (2015).

Miyazaki, S., Imai, T., Igo, Y., and Otsuka, K., "Effect of cyclic deformation on the pseudoelasticity characteristics of Ti-Ni alloys." Metallurgical transactions A, **17** (1), 115-120 (1986).

Neuking, K., Abu-Zarifa, A., Youcheu-Kemtchou, S., and Eggeler, G., "Polymer/NiTi-composites: Fundamental Aspects, Processing and Properties", Advanced Engineering Materials, **7** (11), 1014-1023 (2005).

Ni, Q. Q., Zhang, R. X., Natsuki, T., and Iwamoto, M. (2007), Stiffness and vibration characteristics of SMA/ER3 composites with shape memory alloy short fibers. Composite structures, **79**(4), 501-507.

Nurveren, K., Akdoğan, A., and Huang, W. M., "Evolution of transformation characteristics with heating/cooling rate in NiTi shape memory alloys", Journal of Materials Processing Technology, **196** (1), 129-134 (2008).

Oberaigner, E. R., and Fischlschweiger, M., "A statistical mechanics approach describing martensitic phase transformation", Mechanics of Materials, **43** (9), 467-475 (2011).

Onishi, H., Oyama, M., Soma, T., Kubo, M., Kirimoto, H., Murakami, H., Kameyama, S., "Neuromagnetic activation of primary and secondary somatosensory cortex following tactile-on and tactile-off stimulation" Clinical Neurophysiology, **121**, 588-593 (2010).

Otsuka, K., Saxena, A., Deng, J., and Ren, X., "Mechanism of shape memory effect in martensitic alloys: an assessment" Philosophical Magazine, **91**, 4514-4535 (2011).

Piedboeuf, M.C., Gauvin, R., and Thomas, M., "Damping behaviour of shape memory alloys: Strain amplitude, frequency and temperature effects", Journal of Sound and Vibration, **214** (5), 885-901 (1998).

Pittacio, S., Viscuso, S., "Shape Memory Actuators for Medical Rehabilitation and Neuroscience" Smart Actuation and Sensing Systems-Recent Advances and Future Challenges, InTech, 83-120 (2012)

Raghavan, J., Bartkiewicz, T., Boyko, S., Kupriyanov, M., Rajapakse, N., and Yu, B. (2010), Damping, tensile, and impact properties of superelastic shape memory alloy (SMA) fiber-reinforced polymer composites. Composites Part B: Engineering, **41**(3), 214-222.

Roy, D., Buravalla, V., Mangalgiri, P. D., Allegavi, S., and Ramamurty, U., "Mechanical characterization of NiTi SMA wires using a dynamic mechanical analyzer", Materials Science and Engineering: A, **494** (1), 429-435 (2008).

Sadjadpour, A., Rittel, D., Ravichandran, G., and Bhattacharya, K. "A model coupling plasticity and phase transformation with application to dynamic shear deformation of iron." Mechanics of Materials, **80**, 255-263 (2015).

Schetky, L. M., "Shape Memory Alloys", Scientific American, **241** (5), 74-82 (1979).

Schmoller, K. M., and Bausch, A. R., "Similar nonlinear mechanical responses in hard and soft materials." Nature materials, **12** (4), 278-281 (2013).

Shaw, G. A., Stone, D. S., Johnson, A. D., Ellis, A. B., and Crone, W. C., "Shape memory effect in nanoindentation of nickel–titanium thin films", Applied Physics Letters, **83** (2), 257-259 (2003).

Shaw, J.A., and Kyriakides, S., "Thermomechanical aspects of NiTi." Journal of the Mechanics and Physics of Solids, **43** (8), 1243-1281 (1995).

Sittner, P., Landa, M., Lukas, P., and Novak, V., "R-Phase transformation phenomena in thermomechanically loaded NiTi polycrystals", Mechanics of Materials, **38**, 475-492 (2006).

Sugimoto, K., and Nakaniwa, M., "Change in Young's modulus associated with martensitic transformation in shape memory alloys", Materials Science Forum, **327**, 363-366 (2000).

Van Humbeeck, J., "Shape memory alloys: a material and a technology.", Advanced Engineering Materials, **3** (11), 837-850 (2001).

Van Humbeeck, J., "Damping capacity of thermoelastic martensite in shape memory alloys" Journal of Alloys and Compounds, **355**, 58-64 (2003).

Waitz, T., Antretter, T., Fischer, F.D., Simha, N.K., and Karnthaler, H.P., "Size effects on the martensitic phase transformation of NiTi nanograins", Journal of Mechanics and Physics of Solids, **55**, 419-444 (2007).

Wang, J., and H. Sehitoglu. "Martensite Modulus Dilemma in Monoclinic NiTi-Theory and Experiments", International Journal of Plasticity, **61**, 17-31 (2014).

Wei, Z. G., Sandstrom, R., and Miyazaki, S., "Shape-memory materials and hybrid composites for smart systems: Part I Shape-memory materials", Journal of Materials Science, **33** (15), 3743-3762 (1998a).

Wei, Z. G., Sandstrom, R., and Miyazaki, S., "Shape memory materials and hybrid composites for smart systems: Part II Shape-memory hybrid composites", Journal of Materials Science, **33** (15), 3763-3783 (1998b).

Yastrebov, V. A., Fischlschweiger, M., Cailletaud, G., and Antretter, T., "The role of phase interface energy in martensitic transformations: a lattice Monte-Carlo simulation", Mechanics Research Communications, **56**, 37-41 (2014).

Young, S., Jin, D-H., Chung, D-T, Kim, J-I, Kim, J-I, Nam, and T-H., "Dynamic impact behavior of Ni-rich and Ti-rich shape memory alloys", Materials Research Bulletin, **48**, 5121-5124 (2013).

Zheng, Y., Cui, L., and Schrooten, J., "Temperature memory effect of a nickel–titanium shape memory alloy", Applied Physics Letters, **84** (1), 31-33 (2004).